\documentclass[preprint,aps,floats,nofootinbib,amssymb,superscriptaddress]{revtex4}
\usepackage[utf8]{inputenc}
\usepackage[a4paper]{geometry}
\usepackage{parskip}
\usepackage{amsmath}
\usepackage{amssymb}
\usepackage[labelfont=bf,font=small]{caption}
\usepackage{subcaption}
\usepackage[T1]{fontenc}
\usepackage{graphicx}
\usepackage{hyperref}
\hypersetup{
    colorlinks=true,
    linkcolor=black,
    urlcolor=blue,
    citecolor=black,
}
\begin{document}

%-----------------------------------
% Preprint numbers
%-----------------------------------
%\preprint{ICCUB-}

%-----------------------------------
% Title
%-----------------------------------
\title{\vspace*{1.in}Effect of the cosmological parameters on gravitational waves: general analysis}

%-----------------------------------
% Authors
%-----------------------------------
\author{Dom\`enec  Espriu\footnote{espriu@icc.ub.edu}
  and Marc Rodoreda\footnote{mrodoreda@icc.ub.edu}}
\affiliation{Departament de F\'isica Qu\`antica i Astrof\'isica\, and
Institut de Ci\`encies del Cosmos (ICCUB), \\
Universitat de Barcelona, Mart\'i Franqu\`es 1, 08028 Barcelona, Spain}
\vspace*{2cm}

%-----------------------------------
% Abstract
%-----------------------------------
\begin{abstract}
  Some time ago it was pointed out that the presence of cosmological components could affect the propagation 
of gravitational waves (GW) beyond the usual cosmological redshift and that such effects might be observable 
in pulsar timing arrays. These analyses were done
at leading order in the Hubble constant $H_0$, which is proportional to $\Lambda^{\frac12}$ and 
$\rho_i^{\frac12}$ ($\rho_i$ being the various cosmological fluid densities). In this work, we study in detail 
the propagation of metric perturbations on a Schwarzschild-de Sitter (SdS) background, close to the place 
where GW are produced, and obtain solutions that incorporate corrections linear in
$\rho_i$ and $\Lambda$. At the next-to-leading order the corrections do not appear in the form of $H_0$ 
thus lifting the degeneracy among the various cosmological components. We also determine the leading corrections 
proportional to the mass of the final object; they are very small for the distances considered in pulsar timing 
arrays but may be of relevance in other cases. When transformed into comoving coordinates, the ones used in 
cosmological measurements, this SdS solution does satisfy the perturbation equations
in a Friedmann-Lemaître-Robertson-Walker metric up to and including  $\Lambda^{\frac32}$ terms. This analysis 
is then extended to the other cosmological fluids, allowing us to consider GW sources in the Gpc range. Finally, 
we investigate the influence of these corrections in pulsar timing arrays observations.
\end{abstract}

\maketitle

%%%%%%%%%%%%%%%%%%%%%%%%%%%%%%%%%%%%%%%%%%%%%%%%%%%%%%%%%

\section{Introduction}
Some time ago it was realized \cite{BEP,Espriu2013,Espriu2014} that
the presence of a cosmological constant had an effect on the propagation
of gravitational waves (GW) beyond the modification of the effective frequency
due to the redshift induced by the acceleration of the Universe -
the only effect that is usually taken into account. While this effect
is very small, it was found that it could possibly have observational
consequences.

Later on, the analysis was extended in order to include the effect of the various
cosmological parameters \cite{AEG,EGR,Alfaro}, in particular the matter density $\rho_{\text{dust}}$.
Like in the case when only the dark energy density $\rho_\Lambda$ was included,
the effect led to different corrections in the frequency and in the wave number.
The effective frequency agrees with the usual redshifted one, as expected.

This effect that, as mentioned, has been largely unnoticed does not
have any implications for interferometric experiments
such as LIGO or Virgo \cite{LIGOVirgoFirst,LIGO,Virgo} that depend only on the
GW frequency, but it may be relevant and indeed observable in pulsar timing
arrays (PTA) \cite{Romani1989,FosterBecker1990} where the optical path is much longer 
and therefore sensitive to modifications in the wave number vector.

Still, the effect being roughly proportional to $H_0 L$ (where $L$ is a
characteristic galactic distance and $H_0$ is the present value of the Hubble
constant) is certainly very small
and it is only through a fortuitous combination of various quantities that it is
potentially observable in PTA for GW originating from binary mergers of
very massive black holes, of the order of $10^6$ solar masses at distances
of a few hundreds of Mpc. Although most such mergers take place at further
distances \cite{Supermassive} it was suspected that the modifications should also be visible
for mergers at distances in the Gpc range.

A definite conclusion could not be reached, however, as we were able to solve
the wave equation describing GW propagation in the cosmological medium
only at order $\sqrt{\Lambda}$, or for that matter at order $H_0$ \cite{EGR}. In fact,
at this order of approximation, the various cosmological densities entered only
in the combination forming $H_0$.

In this article we remove these limitations and we are able to provide a
detailed solution up to order $\Lambda^\frac{3}{2}$. This is enough to ensure
the validity of the conclusions for mergers originating in the Gpc range, in fact
almost up to the confines of the visible universe. Interestingly, going beyond
the leading $\sqrt\Lambda$ order (or, equivalently, $H_0$) removes the degeneracy
of the cosmological parameters and $\rho_\Lambda$ and $\rho_{\text{dust}}$ appears in
various combinations.

Of course this is not to say that a potential observation of the effect
at closer distances is not interesting. On the contrary, as emphasized in
\cite{BEP}, this observation could provide a `local' measurement of the
cosmological constant -something interesting in itself. The issue is of course
also related to the ongoing controversy regarding the value of the Hubble constant \cite{Plank,Wong,Verde}.

In addition, we are also able to quantify the possible influence of the gravitational
mass of the source on the wave propagation. As expected, this turns out to be
minute and irrelevant for cosmological distances, but it might be of interest
in other situations, provided that spherical symmetry is still valid to
describe the physical situation.

Just to set the right frame of mind, it is convenient to state outright
the physical cause of the effect. Let us assume that, as a first approximation,
a GW can be described at large distances (but still close to the source) 
by a simple trigonometric function of the form $\frac{1}{r}\cos \omega (t-r)$. In
this formula $t$ and $r$ are the time and radial distance in spherically symmetric
coordinates and for the leading harmonic $\omega$ is equal to twice the orbital frequency
of the collapsing system.  The effect arises because in the presence
of the various cosmological components the universe is not asymptotically flat and the coordinates
$(t,r)$ differ from the ones where observations are made, the
Friedmann-Lemaître-Robertson-Walker comoving system of coordinates $(T,R)$. The
relation is non-trivial and this is the explanation of the fact that wave number
and frequency differ.

General Relativity rules that the spacetime metric reacts to the presence of energy and momentum
according to the Einstein field equations
\begin{equation}\label{Eq: Einstein equations}
R_{\mu \nu} - \frac{1}{2} \, R \, g_{\mu \nu} - \Lambda \, g_{\mu \nu} = \kappa \, T_{\mu \nu},
\end{equation}
where $R_{\mu \nu}$ and $R$ are the Ricci tensor and scalar curvature, $T_{\mu \nu}$ is the energy-momentum 
tensor and $\kappa = 8 \pi G/c^4$. In this paper we will use the $(+---)$ signature convention and 
natural units $c=1$. We have also explicitly included the cosmological constant $\Lambda$ term.

In order to determine the propagation of GW we have to
consider small perturbations around a background metric $\tilde{g}_{\mu \nu}$
\begin{equation}
g_{\mu \nu} = \tilde{g}_{\mu \nu} + h_{\mu \nu}, \qquad |h_{\mu \nu}| \ll 1\ 
\end{equation}
and the linearized vacuum Einstein equations at first order on the perturbation $h_{\mu \nu}$ then read
\begin{equation}
G_{\mu\nu}(\tilde{g}+h) = G_{\mu\nu}(\tilde{g}) + 
\frac{\delta G_{\mu\nu}}{\delta g_{\alpha \beta}}\, \biggl\rvert_{\tilde{g}}\ h_{\alpha\beta} + \ .\ .\ .\ 
= \Lambda \tilde{g}_{\mu \nu} + \Lambda h_{\mu \nu},
\end{equation}
where $G_{\mu \nu}$ is the Einstein tensor. Clearly, Einstein field equations are satisfied for 
the unperturbed metric, $G_{\mu\nu}(\tilde{g}) = \Lambda \tilde{g}_{\mu \nu}$. In 
order to avoid redundancies under coordinate transformations, it is mandatory to choose a gauge. 
Even though there is freedom in the gauge choice, it is convenient to choose coordinates where the 
perturbation is purely spatial $h_{\mu 0} = 0$, transverse and traceless, known as the TT-gauge. 
For a wave propagating in the radial direction, transversality implies that the only non-vanishing 
components of the purely spatial metric perturbation are the angular ones, i.e. 
$h_{\theta \theta}, h_{\theta \phi}, h_{\phi \theta}, h_{\phi \phi}$. Moreover, due to the symmetry of 
the metric tensor, $h_{\theta \phi} = h_{\phi \theta}$. Finally, the traceless condition establishes a 
relation between $h_{\theta \theta}$ and $h_{\phi \phi}$
\begin{equation}
    h = g^{\mu \nu} \, h_{\mu \nu} = 0 \qquad \Rightarrow \qquad h_{\phi \phi} = -\sin^2 \theta \, h_{\theta \theta}.
\end{equation}
The previous considerations apply to any coordinate system displaying rotational symmetry. In
this work two different coordinate systems will be of interest to us: Schwarzschild-de Sitter (SdS) and
Friedman-Lemaitre-Robertson-Walker (FLRW).

\section{Perturbations in Schwarzschild-de Sitter}\label{Section:SdS}
We will first consider the Schwarzschild-de Sitter (SdS) metric
\begin{equation}\label{Eq: SdS metric}
ds^{2}=\left(1-\frac{2 G M}{r}-\frac{\Lambda}{3} r^{2}\right) dt^{2}-
\left(1-\frac{2 G M}{r}-\frac{\Lambda}{3} r^{2}\right)^{-1} dr^{2}-r^{2} d \Omega^{2}.
\end{equation}
This metric possesses spherical symmetry and describes a background consisting of a mass $M$
in a universe endowed with a cosmological constant $\Lambda$. It is the background 'seen' by 
a gravitational wave close to its source, but sufficiently far ($r \gg r_S$) from it so that
spherical symmetry can be considered to hold at least approximately.

This background leads to the following equation of motion for the non-zero components of the metric perturbation.
The equation is the same for $h_{\theta \theta}, h_{\theta \phi}, h_{\phi \theta}$ and $h_{\phi \phi}$ in 
the TT gauge
\begin{align}\label{Eq: h spherical equation}
-\frac{1}{2} \frac{1}{f} \, \ddot{h}_{\mu \nu}
+ \frac{1}{2} f \, h''_{\mu \nu}
+ \left(\frac{1}{2} \, f' - \frac{f}{r} \right) \, h'_{\mu \nu} 
- \left(2 \frac{f'}{r} - \frac{f}{r^2} + \frac{1}{2} \, f'' \right) \, h_{\mu \nu} 
+ \mathcal{O}(h^2) = \Lambda h_{\mu \nu}
\end{align}
where $f(r)=1-\frac{2 G M}{r}-\frac{\Lambda}{3} r^{2}$ is the $g_{tt}$ metric component, a dot $\dot{h}_{\mu \nu}$ 
stands for derivative with respect to time, primes $h'_{\mu \nu} $ represent radial coordinate derivatives 
and $\{\mu, \nu\} = \{\theta, \phi\}$.

Since we observe gravitational waves emitted by very distant sources, we are interested in plane wave solutions. 
Then a more practical coordinate system is the cartesian set of coordinates $\{x,y,z\}$, where spatial coordinates 
can be chosen such as the metric perturbation travels in the z-direction and the source is located in the x-y plane. 
Again, in the TT-gauge for a purely spatial metric perturbation, transversality implies that the only non-zero 
components are $h_{xx},h_{xy},h_{yx}$ and $h_{yy}$. Furthermore, since gravitational waves sources are at located 
at very large distances, we are interested in the small polar angle limit $\theta \approx 0$. Using the 
transformation law for a rank 2 tensor and these considerations, we end up with the following relations 
\begin{equation}\label{Eq: hxx transformation}
h_{xx} = + \frac{1}{r^2} \cos(2\phi) \, h_{\theta \theta} - \frac{1}{r^2} \frac{\cos \theta}{\sin \theta} \sin(2\phi) \, h_{\theta \phi}
\end{equation}
\begin{equation}\label{Eq: hyy transformation}
h_{yy} = - \frac{1}{r^2} \cos(2\phi) \, h_{\theta \theta} + \frac{1}{r^2} \frac{\cos \theta}{\sin \theta} \sin(2\phi) \, h_{\theta \phi}
\end{equation}
\begin{equation}\label{Eq: hxy transformation}
h_{xy} = + \frac{1}{r^2} \sin(2\phi) \, h_{\theta \theta} + \frac{1}{r^2} \frac{\cos \theta}{\sin \theta} \cos(2\phi) \, h_{\theta \phi}.
\end{equation}

We observe that indeed $h_{xx} = - h_{yy}$, as is required in a traceless gauge. Applying this transformation 
in \eqref{Eq: h spherical equation}, the equations of motion for the non-zero cartesian components become
\begin{equation}\label{Eq: h cartesian equation}
    \ddot{h}_{ij}-\left(f^2 \, h''_{ij} +\left(f f' + \frac{2}{r}f^2\right) \, h'_{ij}\right) = 0,
\end{equation}
where now $\{i,j\} = \{x,y\}$. Notice that the equations of motion are actually simpler using cartesian components, 
where all terms proportional to $h_{ij}$ without derivatives cancel out. As expected, in the minkwoskian limit where 
$f \rightarrow 1$, this equation is reduced to the usual spherical wave equation.

In order to solve this equation, we extract a factor $1/r$ from the metric perturbation, since it is 
expected that the amplitude of gravitational waves decreases with the distance from the source
\begin{equation}
    h_{ij} (t,r) = \frac{p_{ij}(t,r)}{r},
\end{equation}
and equation \eqref{Eq: h cartesian equation} takes the following form
\begin{equation}
    \ddot{p}_{ij}-\left(f^2 \, p''_{ij} + f f' p'_{ij}\right) + \frac{ff'}{r} \, p_{ij} = 0.
\end{equation}
Finally, defining the tortoise coordinate \cite{Tortoise} as $dr^*=\frac{1}{f}dr$, the above equation is reduced to
\begin{equation}\label{Eq: p}
    \ddot{p}_{ij} - \partial^2_{r^*} \, p_{ij} + V(r) \, p_{ij} = 0,
\end{equation}
which is a wave equation in a potential $V(r)$ defined as
\begin{equation}\label{Eq: Potential}
    V(r) = \frac{f \, f'}{r}=-\frac{2\Lambda}{3}+\frac{2}{9} \Lambda^2 r^2+\frac{2GM}{r^3}
+\frac{2GM\Lambda}{3\,r}-\frac{(2G M)^2}{r^4}. 
\end{equation}

The tortoise coordinate
in SdS spacetime is usually given in terms of the event 
horizon $r_S \approx 2GM$ and cosmological horizon $r_c \approx \sqrt{3/\Lambda}$, which are solutions of $f(r)=0$. 
Additionally, the function $f(r)$ has another zero at $r_0 = - (r_S+r_c)$, which is not a physical horizon. 
Notice that, while $0<r<r_c$, the tortoise coordinate can take values from $0<r^*<\infty$. The surface gravity 
$\kappa_i$ associated with the horizon $r_i$ is also needed, defined as $\kappa_{i}=\frac{1}{2}|d f / d r|_{r=r_{i}}$. 
With these quantities, the tortoise coordinate can be expressed as \cite{Tortoise}
\begin{equation}
    r^*=\frac{1}{2\kappa_s} \, \log \left(\frac{r}{r_S} -1\right)-\frac{1}{2\kappa_c} \, 
\log \left(1-\frac{r}{r_c}\right)+\frac{1}{2\kappa_0} \, \log \left(1-\frac{r}{r_0}\right).
\end{equation}

In order to solve equation \eqref{Eq: p}, the potential in terms of the tortoise coordinate is needed, so we are 
interested in inverting the above relation to obtain $r(r^*)$. This is not easy but we recall that we are interested
in GW coming from very distant sources, so $r \gg r_S$. Then, the cosmological horizon is 
$r_c \approx \sqrt{3/\Lambda}$, $r_0 \approx - r_c$ and $\kappa_c \approx \kappa_0 = \sqrt{\Lambda/3}$. 
Therefore, in this situation
\begin{equation}\label{Eq: Tortoise coord}
    r^* \approx \sqrt{\frac{3}{\Lambda}} \text{Arctanh} \left(\sqrt{\frac{\Lambda}{3}} \, r\right),
\end{equation}
which can be inverted as $r\approx \sqrt{\frac{3}{\Lambda}} \text{Tanh} \left(\sqrt{\frac{\Lambda}{3}} \, r^*\right)$. 
With this result, it is possible to approximate the wave equation potential in terms of the tortoise coordinate as
\begin{equation}\label{Eq: Potential approx}
    V(r^*) \approx -\frac{2\Lambda}{3}+\frac{2\Lambda}{3} \, \text{Tanh}^2 \left(\sqrt{\frac{\Lambda}{3}}r^*\right) 
+\frac{4GM\Lambda}{3\,r^*}-\frac{(4GM)^2 \Lambda}{9\,{r^*}^2}+\frac{2GM}{{r^*}^3}-\frac{(2GM)^2}{{r^*}^4},
\end{equation}
where $\mathcal{O}(M \Lambda^2)$ and higher order terms in $M$ and $\Lambda$ have been neglected.
Notice that the potential tends to zero as $r^* \rightarrow +\infty$, as it is expected from its 
definition \eqref{Eq: Potential} since $f(r)$ vanishes at $r_c$. 

The relative importance of the various terms in the potential depends of course on the masses and distances involved. 
Assuming that the cosmological constant has the currently preferred value $\Lambda=10^{-52} \text{ m}^{-2}$ \cite{LambdaValue}, 
for masses in the range of 10$^6$ solar masses at $r^* = 10^{24}$ m, the dominant term is $-2 \Lambda/3$, which is of 
order $10^{-53} \text{ m}^{-2}$. The next one, involving a hyperbolic tangent, is a $\mathcal{O}(\Lambda^2)$ term and 
equivalent to $10^{-57} \text{ m}^{-2}$ at this distance. Finally, the mass terms are some orders of magnitude smaller 
in this regime, with $2GM/{r^*}^3$ being the leading one contributing with $10^{-63} \text{ m}^{-2}$ to the potential $V(r^*)$.

Making the comparison one should remember that $r$ here
is not the usual comoving coordinate in FLRW and, in fact, the relation is not one-to-one because the
transformation involves the time coordinate (see below).

Extracting a Fourier factor, we can search 
for solutions $p(t,r^*)=u(r^*) \, e^{-i \omega t}$ in (\ref{Eq: p})
\begin{equation}\label{Eq: u}
    -\omega^2 \, u(r^*)- \partial^2_{r*} \, u(r^*) + V(r^*) \, u(r^*) = 0.
\end{equation}

To begin with, we restrict our problem at distances $r_S \ll r \ll r_c$, where terms in \eqref{Eq: Potential approx} 
proportional to $M$ can be neglected. Also, due to the smallness of the cosmological constant value, we are not 
interested in $\mathcal{O}(\Lambda^2)$ terms. Therefore, the dominant term in the wave equation potential is
\begin{equation}
    V(r^*) \approx -\frac{2\Lambda}{3}.
\end{equation}
With this approximation, equation \eqref{Eq: u} can be solved easily and the desired solution of the metric perturbation
reads at large distances 
\begin{equation}\label{Eq: hij solution}
    h^{\text{SdS}}_{ij} (t,r) = \frac{\epsilon_{ij}}{r} \, \cos{ (\omega t - k r^*)},
\end{equation}
where $\epsilon_{ij}$ is the polarization tensor and the wave number is defined as
\begin{equation}\label{Eq: k definition}
    k^2 = \omega^2+\frac{2\Lambda}{3}.
\end{equation}
Notice that this resembles a dispersion relation corresponding to a massive wave. The reader should not
be alarmed by this. This 'mass-like' term is precisely what is needed for gravitons to have only two polarizations.
This issue is discussed in detail in \cite{Novello}, where the authors analysed the propagation of a massive spin-2 field 
in a de Sitter background and showed that the field has only two degrees of freedom when the mass is $m^2_g = -2 \Lambda /3$, 
and five degrees of freedom otherwise (including $m_g = 0$).

It may be interesting to remind the reader that had the term proportional to $\Lambda r^2$ be omitted altogether
in the Schwarzschild-de Sitter metric, certainly a good approximation close to the source given the smallness of the
cosmological constant, the solution would have  of course been proportional 
to $\cos{ (\omega t - k r)}$. After transformation to FLRW coordinates (see next section)
corrections proportional to $\sqrt\Lambda$ would appear.
The dependence of the tortoise coordinate $r^*$ on $\Lambda$ and having $\cos{ (\omega t - k r^*)}$
is therefore instrumental to recover corrections proportional 
to higher powers of $\Lambda$.

\subsection{Mass corrections}
Let us take into account the dominant mass term in the wave equation potential \eqref{Eq: Potential approx}.
The following approximation for the potential $V(r^*)$ is considered
\begin{equation}
    V(r^*) \approx -\frac{2\Lambda}{3}+\frac{2GM}{{r^*}^3}.
\end{equation}
Taking $u(r^*) =F(r^*) \, e^{ikr^*}$ as an ansatz, where $k$ is given by \eqref{Eq: k definition}, the wave 
equation \eqref{Eq: u} can be written as follows
\begin{equation}
    F''(r^*) + 2 i k \, F'(r^*) - \frac{2GM}{{r^*}^3} F(r^*) = 0,
\end{equation}
where $F'(r^*) \equiv dF(r^*)/dr^*$. This equation can be solved by defining the variable $x=1/r^*$, 
which transforms the above equation into a second-order differential equation that admits an 
infinite series solution
\begin{equation}
    x^2 F'' + (2x - 2 i k) \, F' - 2GM x \, F = 0
\end{equation}
where now $F'=dF/dx$. Therefore, we search for solutions of the form
\begin{equation}
    F = \sum^{\infty}_{n=0} a_n x^{n+s}
\end{equation}
and the following relations between $a_n$ coefficients are found
\begin{align}
    &a_0 = a_0, \qquad a_1 = 0 \\
    &n(n-1)a_n + 2n a_n - (2ik)(n+1) a_{n+1} + (-2GM) a_{n-1} = 0. \nonumber
\end{align}
Since $u(r^*)$ has a plane wave behaviour in the $r^* \rightarrow \infty$ limit, we take $a_0=1$. Finally,
\begin{equation}\label{Eq: F solution}
    F(r^*) = 1 + \frac{-2GM}{2(2ik)} \, \frac{1}{{r^*}^2} + \frac{-2GM}{(2ik)^2} \, \frac{1}{{r^*}^3} + ...
\end{equation}

At large distances from the source $r_S \ll r$, all the terms in the series but the first one are negligible, 
so it is fair to approximate $F(r^*) \approx 1$, recovering the result of the previous analysis 
\eqref{Eq: hij solution}. Putting some numbers, for a supermassive black hole of $10^{10}$ solar masses, at 100 Mpc 
the second coefficient of $F(r^*)$ is approximately $10^{-20} \ll 1$ and totally negligible in that situation. 
However, we take note of this correction as it may be relevant in other physical situations.

\section{Perturbations in FLRW}
We now turn to the description of an expanding de Sitter universe in Friedmann–Lemaître–Robertson–Walker metric, 
which incorporates the physical principles of isotropy and homogeneity. It is expressed 
in comoving coordinates $\{T,R\}$ 
\begin{equation}
ds^2 = dT^2 - a(T)^2\left(\frac{dR^2}{1-KR^2}+ R^2 d\Omega^2 \right),
\end{equation}
where $a(T)$ is the scale factor. In this work we consider a spatially flat ($K=0$) universe. As discussed 
in \cite{AEG}, a comoving cosmological observer will not see the functional form \eqref{Eq: hij solution} 
since $T\neq t$ and $R \neq r$. Therefore, our aim is to relate the previous analysis to this coordinate system.

In order to find the corresponding linearized equations of motion for metric perturbations on a background 
FLRW metric, we proceed as in the previous section. In the transverse and traceless TT-gauge, 
for a wave propagating on the radial direction, the only non-zero metric perturbation components are 
$h_{\theta \theta}, h_{\theta \phi}, h_{\phi \theta}$ and $h_{\phi \phi}$. Then, we switch to a cartesian set of 
coordinates $\{X,Y,Z\}$, where it is possible to choose the $Z$-axis as the propagation direction of the wave 
emitted for a very distant source in the $X$-$Y$ plane. In these coordinates, the non-zero metric perturbation 
components are $h_{XX}, h_{XY}, h_{YX}$ and $h_{YY}$, which are related with the angular ones 
by \eqref{Eq: hxx transformation}-\eqref{Eq: hxy transformation}, where now $r$ is replaced by 
the comoving coordinate $R$. 

Finally, the equations of motion for metric perturbations at first order on the FLRW metric in the TT-gauge are
\begin{align}
-\ddot{h}_{i j} + \left( \frac{\dot{a}}{a} \right) \, \dot{h}_{i j} + \frac{1}{a^2} \left(h''_{i j} + 
\frac{2}{R} \, h'_{i j} \right) + 6\left( \frac{\ddot{a}}{a} \right) \, h_{i j} + 
2\left( \frac{\dot{a}}{a} \right)^2 h_{i j}
= 2 \Lambda \, h_{i j},
\end{align}
where now $\{i,j\}=\{X,Y\}$, and dots and primes stand for derivatives with respect to $T$ and $R$, respectively. 

In this section, a vacuum-dominated universe with only a positive cosmological constant is assumed, with scale factor
\begin{equation}\label{Eq2: Scale factor}
    a(T) = a_0 \exp{\sqrt{\frac{\Lambda}{3}} \, T},
\end{equation}
where $a_0=a(T_0)=1$ is taken at the current time. In this situation, the equations of motion become
\begin{equation}\label{Eq2: EquationFLRW}
\Box_{\text{FLRW}} \, h_{ij} \equiv \left[\partial^2_T - \sqrt{\frac{\Lambda}{3}} \, \partial_T - 
\frac{1}{a^2} \left(\partial^2_R + \frac{2}{R} \, \partial_R \right) - 2 \, \frac{\Lambda}{3} \right] h_{i j}= 0,
\end{equation}
where now the $\partial_i$ notation is used for derivatives for the sake of clarity. This equation again reduces to the Minkwoskian 
wave equation in the absence of a cosmological constant, as expected. For a non-vanishing cosmological constant, 
it is clear that a harmonic function of the variables $T,R$  is not at all a solution of these equations.

In \cite{EGR} it was found that a solution was 
\begin{equation}
  h_{ij}^{\text{FLRW}}= \frac{\epsilon'_{ij}}{R} \, \left(1 + \sqrt{\frac{\Lambda}{3}}T \right) \, \cos \left[\omega \, (T-R)
    + \omega \,  \sqrt{\frac{\Lambda}{3}} \left(\frac{1}{2}R^2-TR\right) \right].
\end{equation}
An arbitrary superposition of such solutions with various frequencies $\omega$ would of course be a solution too
up to order $\sqrt\Lambda$. This is in fact the form that a harmonic wave in coordinates $(t,r)$ of
Schwarzschild-de Sitter takes when is transformed into FLRW coordinates using the coordinate transformation
up to order $\sqrt\Lambda$ (see below). However, at order $\Lambda$ and beyond this will not work. The reason of course
is that, as we saw in the previous section, a simple harmonic is not a solution already at order $\Lambda$ 
in Schwarzschild-de Sitter coordinates either. Let us now discuss this point in some more detail.

\subsection{Discussion}
At large distances from the source and assuming \eqref{Eq2: Scale factor} as scale factor, the exact transformation 
between SdS and FLRW coordinates is given by \cite{BEP}
\begin{subequations}
\label{Eq2: Canvi coord TR}
\begin{align}
    t(T,R) &= T - \frac{1}{2}\sqrt{\frac{3}{\Lambda}} \log \left(1- \frac{\Lambda}{3} \, a^2 \, R^2 \right)\approx T 
+ \frac{1}{2} \sqrt{\frac{\Lambda}{3}} \, R^2 + \frac{\Lambda}{3} \, R^2 T + ... \label{Eq2: Canvi coord T}\\
    r(T,R) &=  a(T) \ R \approx R + \sqrt{\frac{\Lambda}{3}} \, RT + \frac{1}{2}\frac{\Lambda}{3} \, RT^2 
+ ...\label{Eq2: Canvi coord R}
\end{align}
\end{subequations}
The transformation omits the presence of the term proportional to $M$. As we have seen in the previous section
this is totally negligible in the present setting.

Gravitational waves produced by two massive objects in orbit around each other 
would be approximately described, far enough from the source, by harmonic functions periodic in time $t$ 
in SdS coordinates
\begin{equation}\label{Eq2: Previous solution}
    h_{ij} (t,r) = \frac{\epsilon_{ij}}{r} \, \cos{\omega(t - r)}.
\end{equation}
As shown in \cite{EGR}, transforming this harmonic function into comoving coordinates using \eqref{Eq2: Canvi coord TR} 
leads to a solution of the FLRW equation \eqref{Eq2: EquationFLRW} at $\mathcal{O}(\sqrt{\Lambda})$ order 
but it is no longer a solution at the next order.

As discussed in the previous section, a perturbation will propagate in the Schwarzschild-de Sitter spacetime 
approximately as \eqref{Eq: hij solution} far from the source and neglecting $\mathcal{O}(\Lambda^2)$ terms. 
The main differences between \eqref{Eq: hij solution} and \eqref{Eq2: Previous solution} are the appearance of 
the tortoise coordinate $r^*$ inside the argument of the cosine and a different wave number, both of which are 
$\mathcal{O}(\Lambda)$ corrections. Using the relations \eqref{Eq: Tortoise coord} for $r^* \rightarrow r$ 
and \eqref{Eq2: Canvi coord TR} for $\{t,r\} \rightarrow \{T.R\}$, the SdS solution can be expressed 
in comoving coordinates as
\begin{align}\label{Eq2: Transformation solutions}
    h^{\text{SdS}}_{ij} (t,r) = \frac{\epsilon_{ij}}{r} \, \cos{ (\omega t - k r^*)} 
\quad \rightarrow \quad h^{\text{FLRW}}_{ij} (T,R) = \frac{\epsilon'_{ij}}{R} \, a(T) \, \cos \Theta(T,R)
\end{align}
where $\epsilon'_{ij}$ is the transformed polarization tensor, $a(T)$ is the scale factor and the 
argument $\Theta(T,R)$ of the cosine is given by
\begin{equation}\label{Eq2: Argument in TR}
\begin{split}
\Theta(T,R) = \omega \left[T-\sqrt{\frac{3}{\Lambda}} \, \frac{1}{2}\log \left(1- \frac{\Lambda}{3} \, 
a^2 R^2 \right)\right] \\
- \sqrt{\omega^2+\frac{2\Lambda}{3}} \sqrt{\frac{3}{\Lambda}} \text{Arctanh} 
\left(\sqrt{\frac{\Lambda}{3}} \, a R\right).
\end{split}
\end{equation}

Now, this functional form of $h^{\text{FLRW}}_{ij}$ should be a solution of the equations of motion obtained 
by considering first-order perturbations on a FLRW background metric \eqref{Eq2: EquationFLRW} at 
$\mathcal{O}(\Lambda)$ order. And indeed it can be checked that it is.  
In fact, it is a solution for the next order also, i.e. $\mathcal{O}(\Lambda^{3/2})$ order. This analysis is 
valid at large distances from the GW source $r_S \ll r$ and well inside the cosmological horizon, 
where $\Lambda a^2 R^2 \ll 1$ and it is reasonable to neglect the leftover term in \eqref{Eq2: EquationFLRW}
\begin{align}
\Box_{\text{FLRW}} \, h_{ij} = \left(-\frac{2\Lambda}{3}+\frac{2\Lambda}{3} \frac{1}{1-\frac{\Lambda}{3} a^2 R^2 } \right) 
\, h_{ij} \sim \mathcal{O}(\Lambda^2 h).
\end{align}

\subsection{Effective frequency and wave number}\label{Section: Effective freq}
In order to have a closer look at the solution written in comoving coordinates \eqref{Eq2: Transformation solutions} 
and its trigonometric argument \eqref{Eq2: Argument in TR}, we expand them in powers of $\Lambda$
\begin{equation}\label{Eq2: Expanded solution}
h^{\text{FLRW}}_{ij} (T,R) = \frac{\epsilon'_{ij}}{R} \, \left(1 + \sqrt{\frac{\Lambda}{3}}T 
+ \frac{1}{2} \frac{\Lambda}{3} T^2 \right) \, \cos \Theta(T,R)
\end{equation}
\begin{equation}\label{Eq2: Expanded argument}
\begin{split}
\Theta(T,R) = \omega \, (T-R) + \omega \,  \sqrt{\frac{\Lambda}{3}} \left(\frac{1}{2}R^2-TR\right) \\
+ \omega \,  \frac{\Lambda}{3} \left(-\frac{1}{3}R^3+R^2T-\frac{1}{2}RT^2 - \frac{R}{\omega^2}\right),
\end{split}
\end{equation}
where higher terms in $\Lambda$ have been neglected. With this expansion, the anharmonic 
behaviour of the wave as seen by a cosmological observer becomes clear. 

For a GW propagating in an expanding universe, physical intuition tells us that its frequency should
be redshifted as
\begin{equation}
    \omega_{\text{eff}} = \frac{\omega}{1+z}.
\end{equation}
At distances $\Lambda R^2 \ll 1$ that we are considering, the cosmological redshift can be approximated 
by the linear redshift-distance relation $z = H_0 R$, where $H_0$ is the Hubble constant. In fact, this result 
is exact for all distances when the Hubble parameter is constant in time \cite{Harrison}, like in the 
present situation with $H_0 = \sqrt{\Lambda/3}$. Therefore, the expected redshifted frequency at $\Lambda$ order is
\begin{align}\label{Freq redshift}
\omega_{\text{eff}} = \omega \, \left(1-\sqrt{\frac{\Lambda}{3}} \, R + \frac{\Lambda}{3} \, R^2 \right).
\end{align}
At $\sqrt{\Lambda}$ order, the redshift correction term on the frequency appears naturally in \eqref{Eq2: Expanded argument}. 
Imposing the previous relation for the frequency to all orders in $\Lambda$ and rearranging terms
in \eqref{Eq2: Expanded argument}, the following effective wave number is found
\begin{equation}\label{Eq2: Effective k}
 k_{\text{eff}} = \omega \left(1- \frac{1}{2} \, \sqrt{\frac{\Lambda}{3}}R+\frac{\Lambda}{3} \,
\left(+\frac{1}{3}R^2+\frac{1}{2}T^2 + \frac{1}{\omega^2}\right)\right)
\end{equation}
and \eqref{Eq2: Expanded solution} can be written as
\begin{equation}
h^{\text{FLRW}}_{ij} (T,R) = \frac{\epsilon'_{ij}}{R} \, \left(1 + \sqrt{\frac{\Lambda}{3}}T 
+ \frac{1}{2} \frac{\Lambda}{3} T^2 \right) \, \cos \left(\omega_{\text{eff}} \, T - k_{\text{eff}} \, R\right).
\end{equation}
So much for the discussion concerning the cosmological constant only.

\section{General background}
In the previous sections, we have studied the propagation of gravitational waves in a vacuum-dominated universe, 
only filled with a cosmological constant $\Lambda$. Although being the dominant part of the energy and 
matter budget of the universe, the $\Lambda$CDM model incorporates also matter (dark or baryonic) 
and radiation. In terms of their density parameter $\Omega_i$ defined by
\begin{equation}
    \Omega_i \equiv \frac{8\pi G \, \rho_i}{H^2},
\end{equation}
where $\rho_i$ is the density of each species and $H$ the Hubble parameter. The present-day values are 
$\Omega_{\Lambda,0} \sim 0.7$, $\Omega_{\text{dust},0} \sim 0.3$, $\Omega_{\text{rad},0} \sim \mathcal{O}(10^{-5})$. 
While it is safe to neglect the presence of radiation and relativistic matter, the contribution of dust 
is of the same order as the cosmological constant one. The Hubble constant $H_0$ is the current value 
of $H$ and is given by
\begin{equation}\label{Eq3: Hubble constant}
    H_0 = \sqrt{\frac{\kappa \rho_{\Lambda}}{3}+\frac{\kappa \rho_{\text{dust,0}}}{3}+\frac{\kappa \rho_{\text{rad,0}}}{3}}.
\end{equation}

The different types of matter and energy of the Universe are considered as perfect fluids with equation 
of state $p_i=w_i \rho_i$ and included in the energy-momentum tensor as
\begin{equation}
T_{\mu \nu} = (\rho + p)U_{\mu}U_{\nu} - p \, g_{\mu \nu},
\end{equation}
where the fluid four-velocity fulfills the normalization condition $g_{\mu \nu} U^{\mu} U^{\nu} = 1$. It is 
important to note that, while the four-velocity in FLRW comoving coordinates is given by $U^{\mu} = (1,0,0,0)$, 
we have $U^t \neq 1, U^r \neq 0$ in SdS static coordinates. The angular components do vanish in both coordinates 
systems, $U^{\theta} = U^{\phi} = 0$. For a detailed discussion, see Appendix B of \cite{AEG}.
Considering radiation, with $w_{\text{rad}} = 1/3$, pressureless matter, $w_{\text{dust}} = 0$, and vacuum energy 
with a negative pressure, $w_{\Lambda} = -1$, and related with the cosmological constant by 
$\rho_{\Lambda} = \Lambda/\kappa$ we have
\begin{align}
T^{(\text{rad})}_{\mu \nu} = \frac{4}{3} \, \rho_{\text{rad}} U_{\mu}U_{\nu} - \frac{1}{3} \, \rho_{\text{rad}} \, 
g_{\mu \nu} \qquad T^{(\text{dust})}_{\mu \nu} = \rho_{\text{dust}} \, U_{\mu}U_{\nu} \qquad T^{(\Lambda)}_{\mu \nu} 
= \rho_{\Lambda} \, g_{\mu \nu}.
\end{align}
With these ingredients, we proceed as in section \ref{Section:SdS} and consider small perturbations $h_{\mu \nu}$ 
around a background metric. The Einstein equations can be expanded up to first order in the perturbation as
\begin{equation}\label{Eq3: Perturbation theory}
G_{\mu\nu}(\tilde{g}+h) = G_{\mu\nu}(\tilde{g}) + 
\frac{\delta G_{\mu\nu}}{\delta g_{\alpha \beta}}\, \biggl\rvert_{\tilde{g}}\ h_{\alpha\beta} + ... 
= \kappa \left( T^{(0)}_{\mu \nu} + T^{(1)}_{\mu \nu} + ...\right)
\end{equation}
where $\kappa = 8 \pi G$. Again, these field equations are satisfied for an unperturbed metric, 
$G_{\mu\nu}(\tilde{g}) = \kappa T^{(0)}_{\mu \nu}$. 

A spherically symmetric coordinate system that describes the de Sitter space and incorporates the presence of 
dust and (eventually) radiation is needed. The following linearized metric deduced in \cite{AEG,Luciano}  satisfies 
these conditions and reduces to the Schwarzschild-de Sitter metric \eqref{Eq: SdS metric} when dust 
and radiation are not present
\begin{equation}\label{Eq3: General metric}
\begin{split}
    ds^2 = \left(1 - \frac{\kappa}{6} \, ( 2 \rho_{\Lambda} - 2 \rho_{\text{rad}} - \rho_{\text{dust}}) \, r^2 \right) \, dt^2 \\
    - \left(1 + \frac{\kappa}{3} \, ( \rho_{\Lambda} + \rho_{\text{rad}} + \rho_{\text{dust}}) \, r^2 \right) \, dr^2 - r^2 \, d\Omega^2.
\end{split}
\end{equation}
In the above expression the mass term $r_S/r$ has been neglected since it is not important at large distances 
$r_S \ll r$ as we discussed before. We perturb around this background \eqref{Eq3: Perturbation theory} 
and as we have done in section \ref{Section:SdS}, we work in the TT-gauge and with purely spatial components 
of the metric perturbation $h_{\mu \nu}$, where the only non-zero ones are 
$h_{\theta \theta}$, $h_{\theta \phi}$, $h_{\phi \theta}$ and $h_{\phi \phi}$, related by 
$h_{\phi \phi} = - \sin^2 \theta \, h_{\theta \theta}$ and $h_{\phi \theta} = h_{\theta \phi}$. With these ingredients, 
the perturbed Einstein equations neglecting $\mathcal{O}(h^2)$ and higher orders are
\begin{align}\label{Eq3: Equations Spherical}
\frac{1}{f}\, \ddot{h}_{\mu \nu}
+\frac{1}{2}\left(\frac{\dot{g}}{f g}-\frac{\dot{f}}{f^2} \right) \, \dot{h}_{\mu \nu}
- \frac{1}{g} \, h''_{\mu \nu} 
- \left(- \frac{2}{r} \frac{1}{g}
- \frac{1}{2} \frac{g'}{g^2}
+ \frac{1}{2} \frac{f'}{f g}\right) h'_{\mu \nu}
- \bigg(\frac{2}{r^2} \frac{1}{g} \\
+\frac{\ddot{g}}{f g} 
-\frac{1}{2}\frac{(\dot{g})^2}{f g^2} 
- \frac{1}{2} \frac{\dot{f}\dot{g}}{f^2 g}
+ \frac{2}{r}\frac{g'}{g^2}
- \frac{2}{r} \frac{f'}{f g}
+ \frac{1}{2} \frac{f' g'}{f g^2}
+ \frac{1}{2} \frac{(f')^2}{f^2 g}
- \frac{f''}{f g} \bigg)\, h_{\mu \nu}
= -2 \kappa \, T^{(1)}_{\mu \nu},
\nonumber
\end{align}
where $\{\mu, \nu\}$ stands for $\{\theta, \phi\}$, dots and primes for time and radial derivatives, respectively, 
and we have defined the $\tilde{g}_{tt}$ and $\tilde{g}_{rr}$ components of the background metric as
\begin{align}\label{Eq3: f definition}
    f(t,r) &= 1 - \frac{\kappa}{6} \, ( 2 \rho_{\Lambda} - 2 \rho_{\text{rad}} - \rho_{\text{dust}}) \, r^2 \\
    g(t,r) &= 1 + \frac{\kappa}{3} \, ( \rho_{\Lambda} + \rho_{\text{rad}} + \rho_{\text{dust}}) \, r^2 \label{Eq3: g definition}.
\end{align}
Before proceeding, an expression for the perturbed energy-momentum tensor is also needed. In this gauge, we are 
only interested in the angular components so, at first order in the perturbation, they are given by
\begin{align}
T^{(1)}_{\mu \nu} = (\rho_{\Lambda} -\frac{1}{3} \rho_{\text{rad}}) \, h_{\mu \nu}.
\end{align}

Now, considering a GW travelling in the z-direction, we express the components of the metric perturbation in 
cartesian coordinates using the same reasoning as in section \ref{Section:SdS} and the 
\eqref{Eq: hxx transformation}-\eqref{Eq: hxy transformation} relations. In this coordinate system, equations 
\eqref{Eq3: Equations Spherical} take the following form
\begin{equation}\label{Eq3: Equations Cartesian}
\begin{split}
\ddot{h}_{ij}+\frac{1}{2}\left(\frac{\dot{g}}{g}-\frac{\dot{f}}{f} \right) \, \dot{h}_{ij}-\frac{f}{g} \, h''_{ij} 
- \left(\frac{2}{r}\frac{f}{g} + \frac{1}{2} \frac{f'}{g} - \frac{1}{2} \frac{fg'}{g^2} \right) \, h'_{ij}
-\bigg( \frac{\ddot{g}}{g}-\frac{1}{2}\frac{(\dot{g})^2}{g^2} \\
- \frac{1}{2} \frac{\dot{f}\dot{g}}{fg}
- \frac{f'}{g} \frac{1}{r} + \frac{f g'}{g^2} \frac{1}{r} + \frac{(f')^2}{2 f g} + \frac{f' g'}{2 g^2} 
- \frac{f''}{g} \bigg) \, h_{ij} 
= - 2\kappa f (\rho_{\Lambda} -\frac{1}{3} \rho_{\text{rad}}) \, h_{ij}
\end{split}
\end{equation}
where now $\{i,j\}$ stands for $\{x, y\}$. These equations are not easy to solve, but we recall that our aim 
is to find a solution at $\mathcal{O}(\rho)$ order, equivalent to the $\Lambda$ order in the previous sections. 
With this purpose in mind, we study the order in the densities $\rho$ of each term in equation 
\eqref{Eq3: Equations Cartesian}. For the time-derivative terms, assuming a constant $\rho_{\Lambda}$, we have
\begin{align}
    \dot{f} &= \frac{\kappa}{6} \, (2 \partial_t \rho_{\text{rad}} + \partial_t \rho_{\text{dust}}) \, r^2 \\
    \dot{g} &= \frac{\kappa}{3} \, (\partial_t \rho_{\text{rad}} + \partial_t \rho_{\text{dust}}) \, r^2,
\end{align}
and from \cite{AEG,Luciano}, the time derivatives of the densities are
\begin{equation}
    \partial_t (\kappa \rho_{\text{dust}}) = - \frac{A}{3-\kappa \rho_{\text{dust}} r^2} \, 
\frac{(\kappa \rho_{\text{dust}})^{4/3}}{t^{1/3}}
\end{equation}
\begin{equation}
    \kappa \, \partial_t \rho_{\text{rad}} = - \frac{B}{3-\kappa \rho_{\text{rad}} r^2} \, 
(\kappa \rho_{\text{rad}})^{3/2},
\end{equation}
where $A=3 \sqrt[3]{6}$ and $B=4 \sqrt{3}$. Then, in the limit $\kappa \rho_i r^2 \ll 1$, we can approximate 
$\dot{f}$ and $\dot{g}$ as
\begin{equation}
    \dot{f}, \dot{g} \sim (\kappa \rho_{\text{rad}})^{1/2} (\kappa \rho_{\text{rad}} \, r^2) 
+ \left(\frac{\kappa \rho_{\text{dust}}}{t}\right)^{1/3} (\kappa \rho_{\text{dust}} \, r^2).
\end{equation}
Therefore, these terms are of higher order in the densities in this distance regime and can be neglected. 
The same reasoning applies for the $\ddot{g}$ term. Then, the remaining terms proportional to the metric 
perturbation $h_{ij}$ at this order are
\begin{equation}
- \left( - \frac{f'}{g} \frac{1}{r} + \frac{f g'}{g^2} \frac{1}{r} + \frac{(f')^2}{2 f g} 
+ \frac{f' g'}{2 g^2} - \frac{f''}{g} \right) \, h_{ij} + 2 \kappa f \left(\rho_{\Lambda} 
-\frac{1}{3} \rho_{\text{rad}} \right) \, h_{ij}.
\end{equation}
Using the $f$ and $g$ definitions \eqref{Eq3: f definition}-\eqref{Eq3: g definition}, we can study these 
terms at linear order with the densities and observe that they vanish, so this combination of terms proportional 
to the metric perturbation is of order $\rho^2$ or higher
\begin{equation}
\begin{split}
\bigg[ - 4 \, \frac{\kappa}{6} \, ( 2 \rho_{\Lambda} - 2 \rho_{\text{rad}} - \rho_{\text{dust}}) 
- 2 \, \frac{\kappa}{3} \, ( \rho_{\Lambda} + \rho_{\text{rad}} + \rho_{\text{dust}}) \\
+ 2 \kappa \left(\rho_{\Lambda} -\frac{1}{3} \rho_{\text{rad}} \right) + \mathcal{O}(\rho^2) \bigg] \, h_{ij}
\sim \mathcal{O}(\rho^2) \, h_{ij}.
\end{split}
\end{equation}

Consequently, the equations of motion for the cartesian components \eqref{Eq3: Equations Cartesian} at this 
order are simplified as
\begin{equation}\label{Eq3: Final equations}
\ddot{h}_{ij}-\frac{f}{g} \, h''_{ij} - \left(\frac{2}{r}\frac{f}{g} + \frac{1}{2} \frac{f'}{g} 
- \frac{1}{2} \frac{fg'}{g^2} \right) \, h'_{ij}
= 0.
\end{equation}
These equations reduce to the analogous ones on the Schwarzschild-de Sitter metric when $g=1/f$, as expected. 
Using the same strategy to solve these equations than in section \ref{Section:SdS}, we factor out the expected 
$1/r$ behaviour of propagating gravitational waves as $h(t,r)=p(t,r)/r$, so
\begin{equation}\label{Eq3: p equation}
    \ddot{p}_{ij}-\frac{f}{g} \, p''_{ij} - \left( \frac{1}{2} \frac{f'}{g} - \frac{1}{2} \frac{fg'}{g^2} \right) \left(p'_{ij} 
- \frac{p}{r}\right) = 0
\end{equation}
We would like now to introduce a generalized tortoise coordinate that simplifies this differential equation, 
analogous to the SdS case. In fact, this coordinate should recover the SdS form, $dr^* = \frac{1}{f}dr$, 
when $g=1/f$. The desired tortoise coordinate is
\begin{equation}\label{Eq3: Tortoise coord general}
    dr^* = \sqrt{\frac{g}{f}} \, dr
\end{equation}
under which \eqref{Eq3: p equation} takes the following form 
\begin{equation}
    \ddot{p}_{ij}- \partial^2_{r^*} \, p_{ij}+ V(r) \, p_{ij} = 0,
\end{equation}
where we have defined the potential $V(r)$ with the following expression and, at first order in the densities 
and in the $\rho r^2 \ll 1$ regime, is given by
\begin{equation}
V(r) = \frac{1}{2} \frac{1}{r}\left( \frac{f'}{g} - \frac{fg'}{g^2} \right) \approx - \frac{\kappa}{3} \, 
( 2 \rho_{\Lambda} + \frac{1}{2} \rho_{\text{dust}}).
\end{equation}
Although integrating the definition \eqref{Eq3: Tortoise coord general} is not as straightforward as in 
the de Sitter space with only a cosmological constant, a solution can be found in terms of elliptic integrals, 
which can be approximated at first order in the densities as
\begin{equation}\label{Eq3: Tortoise approximation}
   r^* \approx r + \frac{1}{6} \, \frac{\kappa}{3} \, ( 2 \rho_{\Lambda} + \frac{1}{2} \rho_{\text{dust}}) \, r^3.
\end{equation}
Therefore, combining the above results it is possible to find the following solution of equations 
\eqref{Eq3: Final equations} at the considered order
\begin{equation}\label{Eq3: hij general solution}
    h_{ij}(t,r) = \frac{\epsilon_{ij}}{r} \cos (\omega t - kr^*)
\end{equation}
with a wave number
\begin{equation}
k^2 = \omega^2 + \frac{\kappa}{3} \, ( 2 \rho_{\Lambda} + \frac{1}{2} \rho_{\text{dust}}).
\end{equation}
This solution is a generalization of expression \eqref{Eq: hij solution} for a universe filled with pressureless 
matter and radiation. Notice that  $\rho_{\text{rad}}$ does not appear at first order in 
the expression of the potential $V(r)$, so it does not appear in the wave number $k$ one either. This is so because 
the radiation coefficients in the $f$ and $g$ definitions \eqref{Eq3: f definition}-\eqref{Eq3: g definition} 
cancel out when added, a behaviour that affects the studied first order. At higher order in the densities some 
radiation density contributions would eventually emerge. For the same reason, $\rho_{\text{rad}}$ does not appear in the 
tortoise coordinate approximation \eqref{Eq3: Tortoise approximation} either.

\section{Comoving coordinates}
We are interested in how GW are seen by a cosmological observer, so our aim is to express 
the above solution \eqref{Eq3: hij general solution} in comoving coordinates. First, the inclusion of other 
cosmological parameters in \eqref{Eq2: Canvi coord T}-\eqref{Eq2: Canvi coord R} is needed, where only a 
cosmological constant was considered.

For simplicity, in the following analysis the radiation density $\rho_{\text{rad}}$ will be neglected, since 
it does not appear at leading order in the above expressions and its contribution in the current observed 
universe budget seems to be some orders of magnitude lower than $\rho_{\Lambda}$ and $\rho_{\text{dust}}$. Then, it 
is possible to obtain an expression for the scale factor by solving the first Friedmann equation, coming from the 
Einstein field equations
\begin{equation}
\left(\frac{\dot{a}}{a}\right)^2 = \frac{\kappa \, \rho_{\Lambda}}{3} + \frac{\kappa \, 
\rho_{\text{d0}}}{3}\left(\frac{a_0}{a} \right)^3,
\end{equation}
where $\rho_{\text{d0}}$ stands for the current value of the non-relativistic matter density. 
The scale factor is given by \cite{AEG}
\begin{equation}
a(T)=a_{0}\left[\sqrt{1+\frac{\rho_{\text{d0}}}{\rho_{\Lambda}}} \sinh \left(\sqrt{3 \kappa \rho_{\Lambda}} \frac{\Delta T}{2}\right)
+\cosh \left(\sqrt{3 \kappa \rho_{\Lambda}} \frac{\Delta T}{2}\right)\right]^{2/3},
\end{equation}
where $\Delta T = T-T_0$ and $a(T_0)=a_0$, taken as 1 as in the previous section. The scale factor for a 
$\Lambda$-dominated universe \eqref{Eq2: Scale factor} is recovered when $\rho_{\text{d0}} \rightarrow 0$. 
Using this result, it is possible to find the transformation between static SdS coordinates and FLRW comoving ones 
that also preserves spherical symmetric, so the angular element $r^2 d\Omega^2$ becomes $a(T)^2R^2 d\Omega^2$. 
Linearization in terms of the densities leads to the results in \cite{AEG}.
\begin{subequations}
\label{Eq3: Canvi general TR}
\begin{align}
t(T,R) &\approx T+\frac{1}{2} \sqrt{\frac{\kappa}{3}( \rho_{\Lambda}+ \rho_{\text{d0}})} \, 
R^{2}+\left(\frac{\kappa \rho_{\Lambda}}{3}+\frac{\kappa \rho_{\text{d0}}}{12}\right) \Delta T R^{2}
+\ldots \label{Eq3: Canvi general T}\\
r(T,R) &\approx R+\sqrt{\frac{\kappa}{3}( \rho_{\Lambda}+ \rho_{\text{d0}})} \, 
\Delta T R+ \frac{\kappa}{12}\left(2 \rho_{\Lambda}- \rho_{\text{d0}}\right)\Delta T^2 R+\ldots \label{Eq3: Canvi general R}
\end{align}
\end{subequations}
In the limit $\rho_{\text{d0}} \rightarrow 0$ we recover the relations \eqref{Eq2: Canvi coord T}-\eqref{Eq2: Canvi coord R}, 
recalling that $\kappa \rho_{\Lambda}=\Lambda$. Moreover, at order $\sqrt{\rho_0}$ all cosmological densities appear in 
such a combination that reproduces the Hubble constant $H_0$ \eqref{Eq3: Hubble constant}, but this is not so when 
the next order is considered, linear with the densities. Consequently, it is possible {\em a priori} to 
distinguish the contribution from the various densities.

By using these relations along with \eqref{Eq3: Tortoise approximation}, we can transform \eqref{Eq3: hij general solution} 
into FLRW comoving coordinates, keeping $\rho_0$ order terms
\begin{equation}\label{Eq3: General solution comoving}
h_{ij} (T,R) = \frac{\epsilon'_{ij}}{R} \left(1+\sqrt{\frac{\kappa}{3}( \rho_{\Lambda}+ \rho_{\text{d0}})} \, 
T + \frac{\kappa}{12}\left(2 \rho_{\Lambda}-\rho_{\text{d0}}\right)T^2 \right) \cos \Theta(T,R)
\end{equation}
where $\epsilon'_{ij}$ is the transformed polarization tensor and this time the trigonometric argument is given by
\begin{gather}\label{Eq3: General argument}
\Theta(T,R) = \omega \, (T-R) + \omega \,  \sqrt{\frac{\kappa}{3}( \rho_{\Lambda} 
+ \rho_{\text{d0}})} \left(\frac{1}{2}R^2-TR\right) \nonumber\\
+ \omega \, \left[-\left(\frac{\kappa \rho_{\Lambda}}{9} + \frac{\kappa \rho_{\text{d0}}}{36}\right)R^3
+\left(\frac{\kappa \rho_{\Lambda}}{3} + \frac{\kappa \rho_{\text{d0}}}{12}\right)R^2T
-\left(\frac{\kappa \rho_{\Lambda}}{6} - \frac{\kappa \rho_{\text{d0}}}{36}\right)RT^2\right] \\
- \frac{1}{\omega} \left(\frac{\kappa \rho_{\Lambda}}{3} + \frac{\kappa \rho_{\text{d0}}}{12}\right)R \nonumber
\end{gather}
These expressions are a generalization of \eqref{Eq2: Expanded solution}-\eqref{Eq2: Expanded argument}, 
where only a cosmological constant was considered. Also, the different combination of  coefficients proportional 
to the densities $\rho_0$ makes clear that it is not possible to write them in terms of $H^2_0$ only, as stated before.

It is also interesting to express the trigonometric argument as $\omega_{\text{eff}} \, T - k_{\text{eff}} \, R$, 
where the effective frequency $\omega_{\text{eff}}$ satisfies the expected cosmological redshift. 
Analogously to what has been done in section \ref{Section: Effective freq}, the GW frequency will be 
redshifted as $z=H_0 R$
\begin{equation}
 \omega_{\text{eff}} = \omega \left(1 - \sqrt{\frac{\kappa}{3}( \rho_{\Lambda}+ \rho_{\text{d0}})} \, 
R + \frac{\kappa}{3}( \rho_{\Lambda}+ \rho_{\text{d0}}) R^2\right)   
\end{equation}
and the remaining terms form the following effective wave number
\begin{gather}
 k_{\text{eff}} = \omega \bigg[1 - \frac{1}{2} \sqrt{\frac{\kappa}{3}( \rho_{\Lambda}+ \rho_{\text{d0}})} \, R 
 + \left(\frac{\kappa \rho_{\Lambda}}{9}+\frac{\kappa \rho_{\text{d0}}}{36} \right) R^2
 + \left(\frac{\kappa \rho_{\Lambda}}{6}-\frac{\kappa \rho_{\text{d0}}}{12} \right) T^2 + \\
 + \frac{\kappa \rho_{\text{d0}}}{4} TR
 + \frac{1}{\omega^2} \left(\frac{\kappa \rho_{\Lambda}}{3}+\frac{\kappa \rho_{\text{d0}}}{12} \right) \bigg], \nonumber
\end{gather}
which agrees with \eqref{Eq2: Effective k}.

\section{Observational consequences}
Let us summarize briefly our findings. In the previous sections we have been able to derive and solve
the differential equation governing the propagation of GW in a universe endowed with a cosmological
constant and matter density up to order $\rho_\Lambda^\frac{3}{2}$ 
for a vacuum dominated universe and up to order $\rho_\Lambda, \rho_{\text{dust}}$ in the general case.
A limitation of the previous results obtained in \cite{EGR}, where these results had been obtained up to order
$\rho_\Lambda^\frac12$ only, was that going beyond a few Mpc was questionable. The new terms obtained allow us to
explore sources in the Gpc range reliably. In this section we will explore the consequences of the
new corrections in the context of PTA. 

%%%%%%%%%%%%%%%%%%%%%%%%%%
\begin{figure}[t]
\centering
\includegraphics[width=0.33\linewidth]{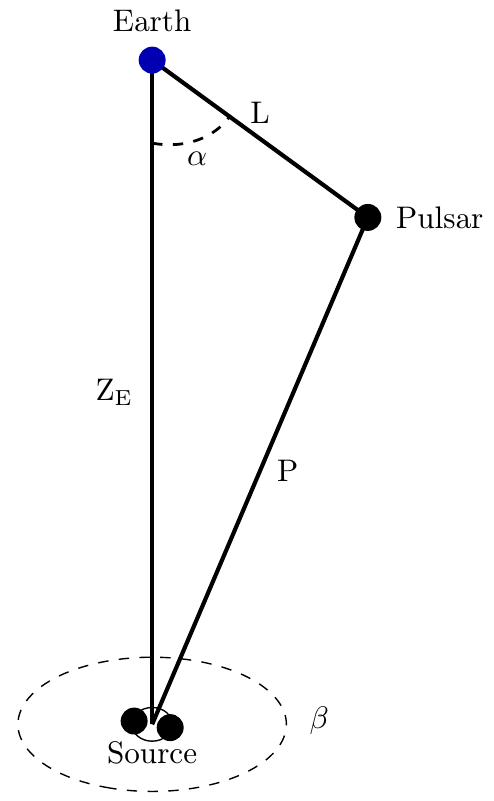}
\caption{Diagram of the relative position of a GW source at a distance $Z_E$ from the Earth and a pulsar
  located at $\vec{P}$ from the source. The angles $\alpha$ and $\beta$ are the polar and azimuthal angles
  of the pulsar with respect to the Earth-source axis.}
\label{Fig: PTA diagram}
\end{figure}
%%%%%%%%%%%%%%%%%%%%%%%%%%

Consider the configuration described in Fig. \ref{Fig: PTA diagram}, where the relative position of a GW source,
a nearby pulsar and the Earth is shown. The pulsar emits electromagnetic pulses with a time-dependent
phase $\phi_0 (T)$, which measured from the Earth reads \cite{Deng}
\begin{equation}
\phi (T) = \phi_0 \left[T - \frac{L}{c} - \tau_0 (T) - \tau_{\text{GW}} (T) \right],
\end{equation}
where we have recovered the speed of light factor $c$, $\tau_0 (T)$ takes into account some corrections on the 
motion of the Earth and the Solar
System and $\tau_{\text{GW}} (T)$ is a timing correction due to the effect of GW. Since a non-zero value modifies
the pulse arrival time, $\tau_{\text{GW}} (T)$ is known as gravitational wave timing residual. It is given by \cite{Deng}
\begin{equation}
\tau_{\text{GW}} (T) = - \frac{1}{2} \, \hat{n}^i \hat{n}^j \, \mathcal{H}_{ij} (T)
\end{equation}
where $\hat{n}$ is a unit vector in the Earth-pulsar direction and $\mathcal{H}_{ij} (T)$ is the integral of the
metric perturbation along the null geodesic from the pulsar to the Earth. The pulsar-Earth path can be parametrized
as $\vec{R}(x) = \vec{P} + L \, (1 +x) \, \hat{n}$, with $x \in [-1, 0]$, so the integral is given by
\begin{equation}\label{Eq4: H integral}
\mathcal{H}_{ij} (T) = \frac{L}{c} \int_{-1}^{0} h_{ij}^{\text{FLRW}}  \left(T_E + \frac{x L}{c}, \vec{R} (x) \right) dx.
\end{equation}
Before proceeding, we make some reasonable approximations. For neighbor pulsars, which are inside our Galaxy, and
GW sources, such as galaxy mergers, we have $L \ll Z_E$. Then, the parametrized path that light follows from the pulsar
to the Earth, in modulus, is $R (x) \approx Z_E + x \, L \cos \alpha$.
Moreover, for a GW propagating in the $Z$ direction, the only non-zero components of the metric perturbation in
the TT-gauge are the spatial $X,Y$ ones. Therefore, we assume for simplicity that the non-zero components of the
transformed polarization tensor are $\epsilon \sim |\epsilon_{ij}|$ for $i, j = X, Y$. Furthermore, we can always
choose a reference plane defined by the position of the Earth, the pulsar and the GW source, so the azimuthal angle
in Fig. \ref{Fig: PTA diagram} can be set to $\beta = 0$. This angle is not important in the timing residual behaviour,
but this is not the case for the polar angle $\alpha$. Finally, the timing residual of the arrival time due to the
passing of gravitational waves reads
\begin{equation}\label{Eq5: Timing residual}
  \tau_{\text{GW}} (T_E, Z_E, L, \alpha, \epsilon, \omega, \rho_i) =
  - \frac{L \epsilon}{2c} \sin^2 \alpha \int_{-1}^{0} h_{ij}^{\text{FLRW}}  \left(T_E + \frac{x L}{c}, Z_E + xL \cos \alpha \right) dx.
\end{equation}
where $h_{ij}^{\text{FLRW}}$ is given by \eqref{Eq2: Expanded solution} for a vacuum-dominated universe and
\eqref{Eq3: General solution comoving} for the $\rho_\Lambda+\rho_\text{dust}$ case.

Our purpose here is simply to assess the relevance of the $\mathcal{O}(\rho)$ corrections with respect to
the $\mathcal{O}(\sqrt\rho)$ corrections previously known.
In order to perform a numerical analysis, we take reasonable values for some parameters appearing in
\eqref{Eq5: Timing residual}. We choose $\epsilon= 1.2 \times 10^9 \text{ m}$, so that $|h| \sim \epsilon / Z \sim 10^{-15}$,
and $\omega = 10^{-8} \text{ rad/s}$, corresponding to ultra-low frequency GW signals, which are values within
the sensibility of PTA projects \cite{Supermassive,Jenet}. We consider one pulsar located at $L = 0.3 \text{ kpc}$ from the Earth.

\begin{figure}[t]
\begin{subfigure}{0.49\textwidth}
  \centering
  \includegraphics[width=\linewidth]{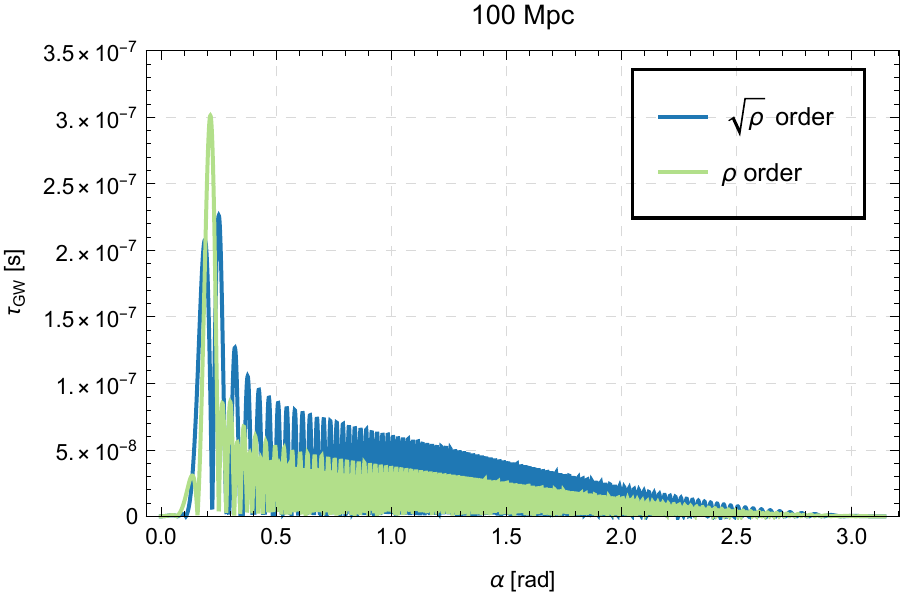}
  \label{Fig: 100Mpc}
\end{subfigure}
\begin{subfigure}{0.49\textwidth}
  \centering
  \includegraphics[width=\linewidth]{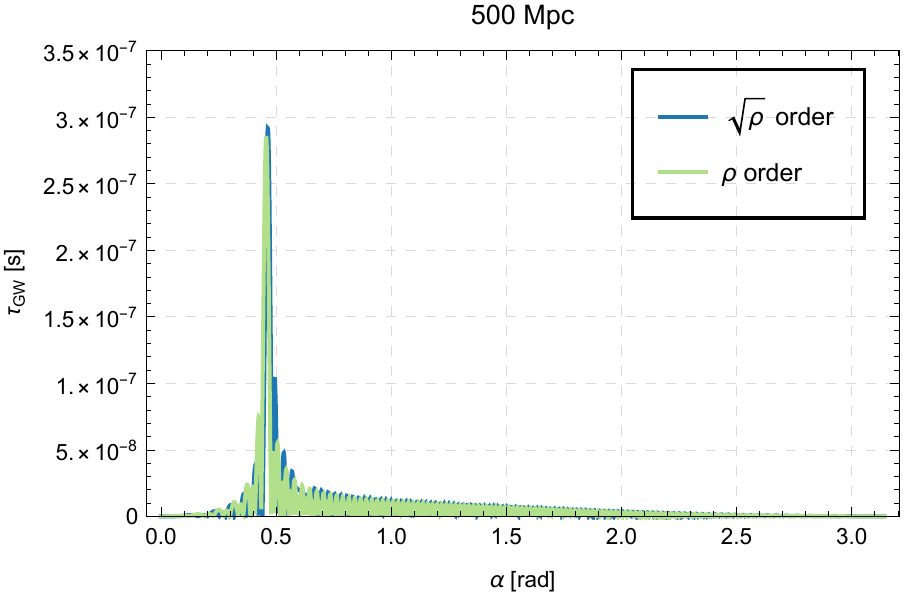}
  \label{Fig: 500Mpc}
\end{subfigure}
\begin{subfigure}{0.49\textwidth}
  \centering
  \includegraphics[width=\linewidth]{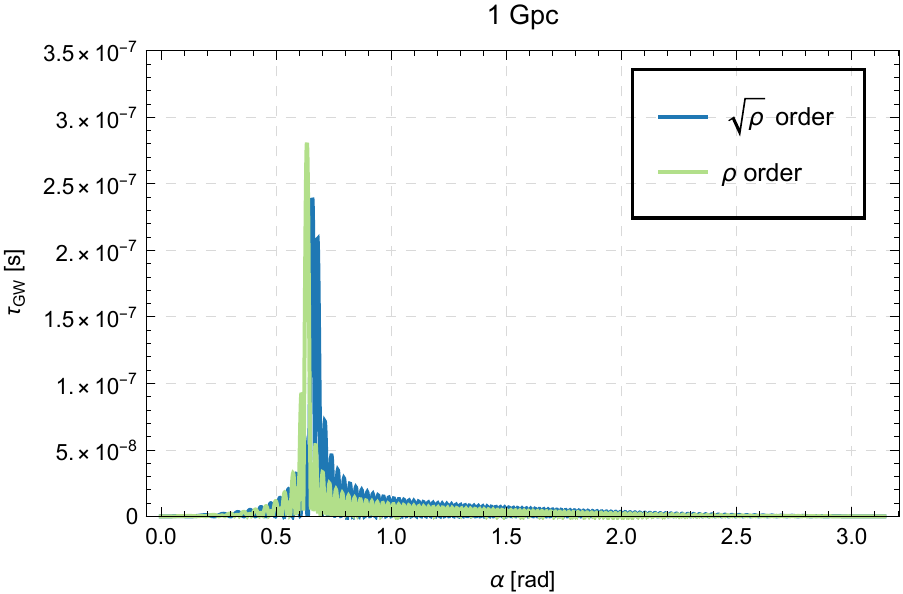}
  \label{Fig: 1Gpc}
\end{subfigure}
\begin{subfigure}{0.49\textwidth}
  \centering
  \includegraphics[width=\linewidth]{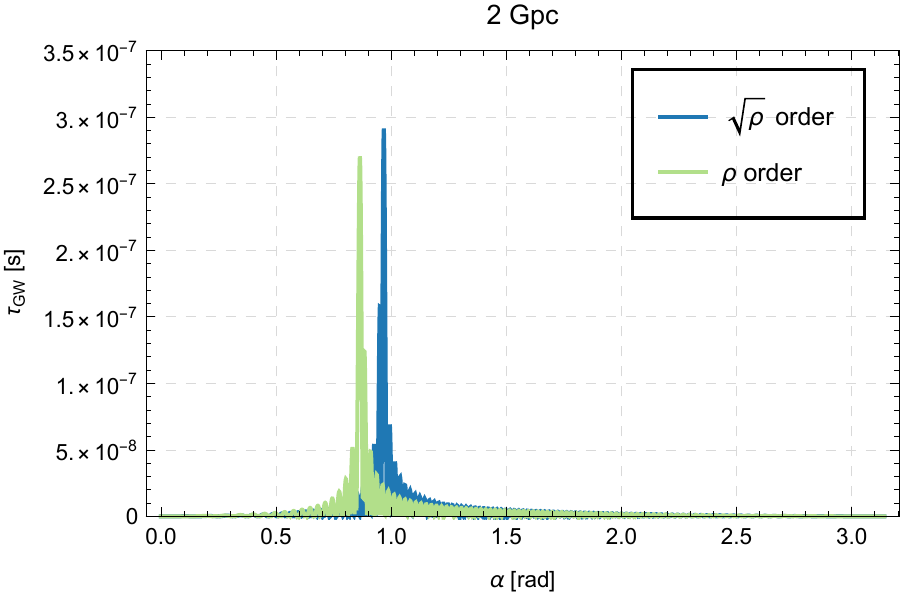}
  \label{Fig: 2Gp}
\end{subfigure}
\caption{Comparison of the absolute timing residual $|\tau_{\text{GW}}|$ between the $\sqrt{\rho}$ (blue) and $\rho$ (green) order solutions for sources located at 100 Mpc, 500 Mpc, 1 Gpc and 2 Gpc. The figures are symmetrical for $\pi \leq \alpha \leq 2\pi$.}
\label{Fig: Sqrt vs Rho}
\end{figure}

In \cite{Espriu2013,Espriu2014,AEG,EGR,Alfaro} it was shown that the presence of a non-zero cosmological constant and 
other cosmological fluids
could affect the timing residuals. In Fig. \ref{Fig: Sqrt vs Rho} we compare the resulting timing residual
for the already known solution $h_{ij}$ at $\sqrt{\rho}$ order with the inclusion of $\rho$ order corrections,
given by \eqref{Eq3: General solution comoving}, for a universe filled with dark energy and dark matter.

A remarkable feature of these plots is an important enhancement of the signal for a particular value of the
angle $\alpha_m$, where the timing residual reaches its maximum. The position of this peak depends on the distance
to the source $Z_E$, which occurs at larger angles for further sources. While the peak appears at similar
angular positions for both cases, the corrections linear with the densities $\rho$ allow us to safely
explore remoter sources in the Gpc region where most mergers are expected to occur. The 
peak $\alpha_m$ is at slightly lower values when the $\rho$ corrections are included, particularly for
very distant sources. The main conclusion
of the present study is that it is actually viable to seek for the effect of very massive black hole
mergers at any distance.

%%%%%%%%%%%%%%%%%%%%%%%%%%%%%%%%%%%%
%%%%%%%%%%%%%%%%%%%%%%%%%%%%%%%%%%%%

\newpage
\section{Conclusions}

It was already known that a harmonic function like \eqref{Eq2: Previous solution} in $\{t,r\}$ coordinates,
which describes well gravitational waves far away from their source, is a solution of the equations of motion for
perturbations on a FLRW background metric \eqref{Eq2: EquationFLRW} only up to $\sqrt{\Lambda}$ order when
transformed into comoving coordinates $\{T,R\}$. In order to go to the next order, we have studied metric
perturbations on the SdS metric in section \ref{Section:SdS} and obtained $h^{\text{SdS}}(t,r)$ \eqref{Eq: hij solution},
which includes $\mathcal{O}(\Lambda)$ corrections inside the argument of the cosine. This functional form,
transformed into comoving coordinates, does satisfy the FLRW perturbation equations in the TT-gauge up to $\Lambda^2$ terms.
The previous discussion is valid well inside the cosmological horizon.

In addition, we have extended the analysis to include all other cosmological fluids up to order $\rho$. 
We provide explicit formulae
for the effective wave number. This result is non-trivial. Furthermore, it is found that beyond the leading order
the densities appear in combinations other than $H_0$. This potentially removes degeneracies in what concerns the
propagation of gravitational waves in a cosmological background. These results support the conclusions put forward 
in previous works \cite{Espriu2013,Espriu2014,AEG,EGR,Alfaro} concerning the possible measurement of the cosmological 
parameters in PTA observations. In fact, as emphasized
e.g. in \cite{AEG}, this effect may {\em facilitate} a positive detection of GW in PTA.

In passing, we have derived a number of interesting results, such as the dependence of the propagation equation on the
final mass of the merger producing the gravitational waves, which is minute but possibly conceptually relevant.

\section*{Acknowledgements}
We would like to thank our collaborators J. Alfaro, J. Bernabeu, L. Gabbanelli and D. Puigdom\`enech.  
This research is partly supported by the Ministerio de Ciencia e Inovaci\'on under research grants
PID2019-105614GB-C21, CEX2019-000918-M of ICCUB (Unidad de Excelencia Mar\'\i a de Maeztu),  
and by grant 2017SGR0929 (Generalitat de Catalunya).

\end{document}